\documentclass{jpsj2}
\title{Dynamical Mean-Field Theory and Its Applications to Real
Materials}

\author{D.\ Vollhardt$^1$, K.\ Held$^2$, G.\ Keller$^1$, R.\ Bulla$^1$,
Th.~Pruschke$^{3}$,  I.~A.~Nekrasov$^4$, and V.~I.~Anisimov$^4$ }

\inst{ $^1$Theoretische Physik III, Elektronische Korrelationen
und Magnetismus, Institut f\"ur Physik, Universit\"at Augsburg,
86135 Augsburg, Germany\\
$^2$Max-Planck-Institut f\"ur
Festk\"orperforschung, 70569 Stuttgart, Germany\\
$^3$Institut f\"ur Theoretische Physik, Universit\"at G\"ottingen,
Friedrich-Hund-Platz~1,
D-37077 G\"ottingen, Germany\\
$^4$Institute of Metal Physics, S. Kovalevskoj Str. 18,
Ekaterinburg GSP-170, 620219 Russia}

\abst{Dynamical mean-field theory (DMFT) is a non-perturbative
technique for the investigation of correlated electron systems.
Its combination with the local density approximation (LDA) has
recently led to a material-specific computational scheme for the
\emph{ab initio} investigation of correlated electron materials.
The set-up of this approach and its application to materials such
as (Sr,Ca)VO$_3$, V$_{2}$O$_{3}$, and Cerium is discussed. The
calculated spectra are compared with the spectroscopically
measured electronic excitation spectra. The surprising similarity
between the spectra of the single-impurity Anderson model and of
correlated \emph{bulk} materials is also addressed.}

\kword{Strongly correlated electron systems, Mott-Hubbard
metal-insulator transition, dynamical mean-field theory, ab initio
computational approaches}

\begin{document}
\maketitle

\section{Introduction}
In many materials with partially filled $d$ and $f$ electron
shells, such as the transition metals Ti, V, Fe and their oxides
or rare--earth metals such as Ce, electrons occupy narrow
orbitals. This spatial confinement enhances the effect of the
Coulomb interaction between the electrons, making them ``strongly
correlated''.\cite {tokura} The interplay between the spin, charge
and orbital degrees of freedom of the correlated $d$ and $f$
electrons and the lattice degrees of freedom leads to a multitude
of unusual ordering phenomena at low temperatures. Consequently,
strongly correlated electron systems are often exceedingly
sensitive to small changes in the temperature, pressure, magnetic
field, doping, and other control parameters. This results, for
example, in large changes of the resistivity across
metal-insulator transitions,
of the volume across phase transitions, and of the effective
electronic masses. Electronic correlations are also essential for
an understanding of high temperature superconductivity. These
properties cannot be explained within conventional mean-field,
e.g., Hartree-Fock theory, since these theories describe the
interaction only in an average way and in terms of a \emph{static}
mean field.

\section{Dynamical Mean-Field Theory (DMFT)}
During the last few years, our understanding of electronic
correlation effects has considerably improved due to the
development of dynamical mean-field theory
(DMFT)~\cite{DMFT_vollha, Brandt, MH, Vaclav, Georges92,
Jarrell92, vollha93, pruschke, DMFT_georges}; for an introduction
into DMFT and its applications see Ref.\cite{PT}. Within DMFT the
electronic lattice problem is mapped onto a single-impurity
Anderson model with a self-consistency
condition.~\cite{DMFT_georges} This mapping becomes exact in the
limit of large coordination number of the
lattice~\cite{DMFT_vollha} and allows one to investigate the
dynamics of correlated lattice electrons non-perturbatively at all
interaction strengths.

The single-impurity Anderson model \cite{Anderson61} itself
defines one of the fundamental many-body problems whose
investigation during the last 40 years has led to enormous
physical insights and important progress in the development of
theoretical investigation techniques\cite{Hewson93}. In
particular, a new many-body energy scale $T_K$ (the
Kondo\cite{Kondo} temperature) is known to arise in this problem,
which is exponentially small in the limit of vanishing
hybridization. Below this temperature the system can always be
understood as a ``local Fermi liquid'' with strongly renormalized
quasiparticles.\cite{Nozieres74} In particular, the
single-particle spectrum, i.e., the local density of states (DOS),
exhibits a generic three peak structure consisting of two broad
maxima originating from the atomic levels, and a sharp ``Kondo''
or ``Abrikosov-Suhl'' resonance at the Fermi level, of width
$T_K$.

We now know that this characteristic three-peak spectrum of the
single-impurity model is also found in correlated \emph{bulk}
systems, i.e., lattice models and real materials, where the notion
of a ``single impurity'' is not applicable. This surprising fact
is explained by DMFT. Indeed, DMFT is presently the only
theoretical approach which can derive the electronic excitation
spectrum at all energy scales, hence reproducing the incoherent
features at high energies (Hubbard bands),~\cite{HubbardIII}
\emph{and} the coherent quasiparticle behavior at low energies
within the same formalism.~\cite{Gutzwiller, brinkman70}

\section{Single-Particle Spectrum of Correlated Electrons: One-Band Hubbard Model}

\begin{figure}[t]
\begin{center}
\includegraphics[width=8cm]{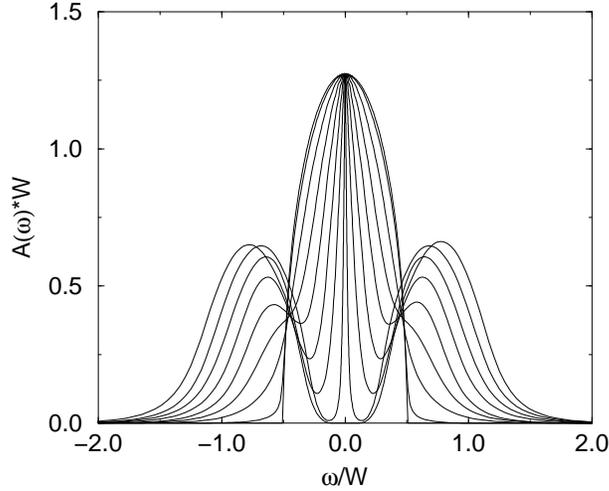}
\end{center}
\caption[]{Evolution of the $T=0$ spectral function of the
one-band Hubbard model with a semi-elliptic (``Bethe'') DOS for
interaction values $U/W=0,0.2,0.4,\ldots,1.6$ ($W$: band width)
calculated with the numerical renormalization group\cite{Bulla}.
At the critical interaction $U_{\rm c2}/W\simeq1.47$ the metallic
solution disappears and the Mott gap opens. The ``Luttinger
pinning'' at $\omega=0$ is clearly observable.}
\label{DOS_pinning}
\end{figure}

Apparently there is a close relation between quantum impurity
physics, with the Kondo problem as its paradigm, and the physics
of correlated electrons on a lattice as exemplified by the Hubbard
model. The connection on the technical side is through the mapping
of a lattice model onto an effective impurity model, as done in
DMFT. As a consequence, well-known features of the impurity model
such as the Kondo resonance reappear in the solution of lattice
models. In particular, the pinning of the density of states (DOS)
at the Fermi level obtained within DMFT\cite{MH} directly
corresponds to the Friedel sum-rule for the single-impurity
Anderson model. This is clearly seen in Fig. \ref{DOS_pinning}
where we show the evolution of the local spectral function
$A(\omega )$ of the DMFT solution for the one-band Hubbard model

\begin{eqnarray}
\hat{H} &=& -t\sum_{{ (ij)},\sigma}
 c_{{ i}\sigma}^{\dagger}c_{{ j}\sigma}^{\phantom{\dagger}}
 + U\sum_{{ i}} n_{{ i}\uparrow}^{\phantom{\dagger}}
         n_{{i}\downarrow}^{\phantom{\dagger}}\label{H_Hub}
\end{eqnarray}
at zero temperature and half filling as a function of local
Coulomb repulsion $U$ in units of the bandwidth $W$ of
non-interacting electrons. Here $i$ denotes the lattice site, $t$
is the amplitude for nearest-neighbor hopping on the lattice, and
$U$ is the local Hubbard repulsion. Magnetic order is assumed to
be suppressed (``frustrated'').
While at small $U$ the system can be described by quasi-particles
whose DOS still resembles that of the free electrons, in the Mott
insulator state the spectrum consists of two separate ``Hubbard
bands'' whose centers are separated approximately by the energy
$U$. The latter originate from ``atomic'' excitations at the
energies $\pm U/2$ broadened by hopping of electrons away from the
atom. At intermediate values of $U$ the spectrum then has a
characteristic three-peak structure as in the single-impurity
Anderson model, which includes both the atomic features (i.e.,
Hubbard bands) and the narrow quasi-particle peak at low
excitation energies, near $\omega=0$.~\cite{DMFT_georges} This
corresponds to a strongly correlated metal. The structure of the
spectrum (lower Hubbard band, quasiparticle peak, upper Hubbard
band) is quite insensitive to the specific form of the DOS of the
non-interacting electrons.

It is important to note that the three-peak spectrum in
Fig.~\ref{DOS_pinning} originates from a lattice model with
\emph{one} type of electrons only. This is in contrast to the
single--impurity Anderson model whose spectrum shows very similar
features, but is due to \emph{two} types of electrons, namely the
localized orbital at the impurity site and the free conduction
band. Therefore the screening of the magnetic moment which gives
rise to the Kondo effect in impurity systems has a different
origin in lattice systems. Namely, as explained by DMFT, the same
sort of electrons provide both the local moments and the electrons
which screen these moments. DMFT also allows one to investigate
the {\em periodic} Anderson model
  in which a lattice of localized electrons hybridizes with an
  uncorrelated conduction band (see, for example, Refs.~\cite{pam}).
  The resulting DOS of the localized electrons again shows the typical
  three-peak structure. There are, however, important differences
  to both the Hubbard model and the single-impurity Anderson model,
  such as the appearance of a hybridization gap at or close to the
  Fermi level.

The vanishing of the quasiparticle peak in the Hubbard model
signals a ``Mott-Hubbard metal-insulator transition''.
This transition between a paramagnetic metal and a paramagnetic
insulator induced by the Coulomb interaction between the electrons
is one of the most famous examples of a cooperative phenomenon
involving electronic correlations. The question concerning the
microscopic origin and nature of this transition poses one of the
fundamental theoretical problems in condensed matter
physics.\cite{Mott,Gebhard} Correlation induced metal-insulator
transitions (MIT) are found, for example, in transition metal
oxides with partially filled bands near the Fermi level such as
V$_{\!2}$O$_{3}$ doped with Cr\cite{McWhR,McWhetal} (see section
5.2). For these systems band theory typically predicts metallic
behavior.

>From a theoretical point of view the Mott transition is a
paradigmatic correlation problem since it focusses on the
competition between kinetic energy and correlation energy of
correlated electrons in the solid. Here DMFT has led to
significant new
insights.\cite{PT,DMFT_georges,Roz99,Joo,Bulla,Bulla01} In
particular, the Mott-transition is found to be of first order at
finite temperatures, being associated with a hysteresis region in
the interaction range $U_{\rm c1}<U<U_{\rm c2}$ where $U_{\rm c
1}$ and $U_{\rm c 2}$ are the values at which the insulating and
metallic solution, respectively, vanishes. The hysteresis region
terminates at a critical point $(U^\ast,T^\ast)$. For temperatures
above $T^\ast$ the transition changes into a smooth crossover from
a bad metal to a bad insulator.

\begin{figure}[t]
\begin{center}\mbox{}
\includegraphics[width=10cm]{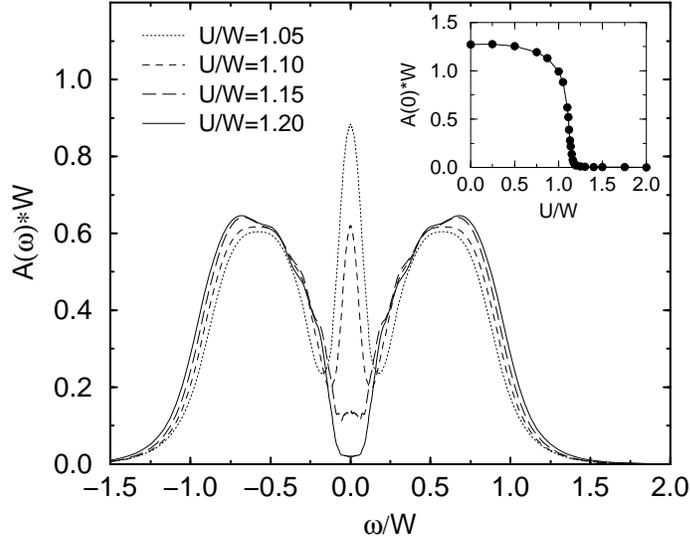}
\end{center}
\caption[]{Spectral function for the half-filled Hubbard model for
various values of $U$ at $T=0.0276W>T^\ast$ (in the crossover
region). The crossover from the metal to the insulator occurs via
a gradual suppression of the quasiparticle peak at $\omega\!=\!0$.
The inset shows the $U$-dependence of $A(\omega\! =\!0)$, in
particular the rapid decrease for $U\approx 1.1W$ [after Bulla,
Costi, and Vollhardt\cite{Bulla01}].} \label{fig:4.1}
\end{figure}

The evolution of the spectral function of the half-filled
frustrated Hubbard model at \emph{finite} temperatures,
$T=0.0276W$, is shown in Fig.~\ref{fig:4.1}. This temperature is
above the temperature of the critical point so that there is no
real transition but a crossover from a metallic-like to an
insulating-like solution.
The height of the quasiparticle peak at the Fermi energy is no
longer fixed at its zero temperature value. This is due to a
finite value of the imaginary part of the self--energy.
The spectral weight of the quasiparticle peak is seen to be
gradually redistributed and shifted to the upper (lower) edge of
the lower (upper) Hubbard band. The inset of Fig.~\ref{fig:4.1}
shows the $U$-dependence of the value of the spectral function at
zero frequency $A(\omega\!=\!0)$. For higher values of $U$ the
spectral density at the Fermi level is still finite and vanishes
only in the limit $U\to\infty$ (or for $T\to 0$, provided that
$U>U_{\rm c2}(T=0)$).

For the insulating phase DMFT predicts the filling of the
Mott-Hubbard gap with increasing temperature. This is due to the
fact that the insulator and the metal are not distinct phases in
the crossover regime, implying that the insulator has a finite
spectral weight at the Fermi level. This behavior has recently
been detected experimentally by photoemission
experiments.\cite{Mo04}

\section{LDA+DMFT}
\label{LDA+DMFT}

Although the Hubbard model is able to explain certain basic
features of the Mott-Hubbard MIT and of the phase diagram of
correlated electrons it cannot explain the physics of real
materials in any detail. Clearly, realistic theories must take
into account the explicit electronic structure of the systems.

Recently, the LDA+DMFT method, a new computational scheme that
merges electronic band structure calculations and the dynamical
mean-field theory, was developed
\cite{Anisimov97a,Zoelfl00,Nekrasov00,psi-k,licht,PT}. Starting
from conventional band structure calculations in the local density
approximation (LDA) the correlations are taken into account by the
Hubbard interaction and a Hund's rule coupling term. The resulting
DMFT equations are solved numerically with a quantum Monte-Carlo
(QMC) algorithm. By construction, LDA+DMFT includes the correct
quasiparticle physics and the corresponding energetics and
reproduces the LDA results in the limit of weak
Coulomb interaction $U$.
More importantly, however, LDA+DMFT correctly describes the
correlation induced dynamics near a Mott-Hubbard MIT and beyond.
Thus, LDA+DMFT is able to account for the correct physics for all
Coulomb interactions and doping levels.

 In the LDA+DMFT approach
\cite{Anisimov97a,psi-k,licht} the LDA band structure is expressed
by a one-particle Hamiltonian $\hat{H}_{\mathrm{LDA}}^{0}$, and is
then supplemented by the local Coulomb repulsion $U$ and Hund's
rule exchange $J$. This leads to a material specific
generalization of the one-band model Hamiltonian (1):

\begin{eqnarray}
\hat{H} &=&\hat{H}_{\mathrm{LDA}}^{0}+{\ U}\sum_{m}\sum_{i}\hat{n}%
_{im\uparrow }\hat{n}_{i m\downarrow } \;+\;\sum_{i, m\neq {m}^{'}, \sigma,%
{\sigma}^{'}}\;(V-\delta _{\sigma {\sigma}^{'}}J)\;\hat{n}_{im\sigma }%
\hat{n}_{im^{'}\sigma^{'}}.  \label{H}
\end{eqnarray}%

\begin{eqnarray}
\hat{H}_{\mathrm{LDA}}^{0} &=& \hat{H}_{\rm{LDA}}
-{\sum_{i}}\sum_{m\sigma} \Delta\epsilon_d \,\hat{n}_{im\sigma}.
\label{Hint}
\end{eqnarray}

Here $m$ and $m^{'}$ enumerate the  three interacting $t_{2g}$
orbitals of the transition metal ion or the $4f$ orbitals in the
case of rare earth elements.
The interaction parameters are related by $%
V=U-2J$ which holds exactly for degenerate orbitals and is a good
approximation for the $t_{2g}$. The actual values for $U$ and $V$
can be obtained from an averaged Coulomb parameter $\bar U$ and
Hund's exchange $J$, which can be calculated by constrained LDA.
The energy term containing $\Delta\epsilon_d$ is a shift of the
one-particle potential of the interacting orbitals. It cancels the
Coulomb contribution to the LDA results, and can be calculated by
constrained LDA.\cite{psi-k}

Within the LDA+DMFT scheme the self-consistency condition
connecting the self-energy $\Sigma $ and the Green function $G$ at
frequency $\omega$ reads: \vspace{-0.5cm}

\begin{eqnarray}
G_{qm,q^{\prime }m^{\prime }}(\omega )=\!\frac{1}{V_{B}}\int{{\
d^{3}}{k}} \!&\left( \left[ \;\omega {\bf 1}+\mu {\bf
1}-H_{\mathrm{LDA}}^{0}
(\mathbf{k}%
) -\Sigma(\omega )\right]^{-1}\right)_{q  m,q^{\prime }m^{\prime
}} .&  \label{Dyson}
\end{eqnarray}

Here, ${\bf 1}$ is the unit matrix, $\mu$ the chemical potential,  $H_{\mathrm{LDA}%
}^{0}(\mathbf{k})$ is the  orbital matrix  of the
 LDA Hamiltonian derived, for example, in a linearized
muffin-tin
orbital (LMTO) basis, $%
\Sigma(\omega)$ denotes the self-energy matrix which is nonzero
only between the interacting orbitals, and $[...]^{-1}$ implies
the inversion of the matrix with elements $n$ (=$qm$), $n^{\prime
}$(=$q^{\prime }m^{\prime }$), where $q$ and $m$ are the indices
of the atom in the primitive cell and of the orbital,
respectively. The integration extends over the Brillouin zone with
volume $V_{B}$ (note that $\hat{H}_{\mathrm{LDA}}^{0}$ may include
additional non-interacting orbitals).

For cubic transition metal oxides
 Eq.\ (\ref{Dyson})
can be simplified  to
\begin{eqnarray}
G(\omega)&\!=\!&G^{0}(\omega-\Sigma (\omega))=\int d\epsilon
\frac{N^{0}(\epsilon )}{\omega-\Sigma (\omega)-\epsilon}
\label{intg}
\end{eqnarray}
if the degenerate $t_{2g}$ orbitals crossing the Fermi level are
well separated from the other orbitals.~\cite{psi-k} For non-cubic
V$_2$O$_3$ the degeneracy is lifted. In this case we employ Eq.\
(\ref{intg}) as an approximation, using different $\Sigma_m
(\omega)$, $N^{0}_m(\epsilon )$ and $G_m(\omega)$ for the three
non-degenerate
 $t_{2g}$ orbitals.

The Hamiltonian (\ref{H}) is solved within the DMFT using standard
quantum Monte-Carlo (QMC) techniques to solve the self-consistency
equations~\cite{QMC}. From the imaginary time QMC Green function
we calculate the physical (real frequency) spectral function with
the maximum entropy method,~\cite{MEM} using the program by
Sandvik.

\section{Single-Particle Spectrum of Correlated Electrons: Real Materials}

Transition metal oxides are an ideal laboratory for the study of
electronic correlations in solids. Among these materials, cubic
perovskites have the simplest crystal structure and thus may be
viewed as a starting point for understanding the electronic
properties of more complex systems. Typically, the $3d$ states in
those materials form comparatively narrow bands with width
$W\!\!\sim \!2\!-\!3\,$~eV, which leads to strong Coulomb
correlations between the electrons. Particularly simple are
transition metal oxides with a 3$d^{1}$ configuration since, among
others, they do not show a complicated multiplet structure.

Photoemission spectra provide a direct experimental tool to study
the electronic structure and spectral properties of electronically
correlated materials. Intensive experimental investigations of
spectral and transport properties of strongly correlated 3$d^{1}$
transition metal oxides started with investigations by Fujimori
\textit{et al.}~\cite{fujimori}. These authors observed a
pronounced lower Hubbard band in the photoemission spectra (PES)
which cannot be explained by conventional band structure theory.

In the following we will employ LDA+DMFT to compute the
$\mathbf{k}$-integrated electronic spectra of two correlated
materials, the 3$d^{1}$ system (Sr,Ca)VO$_3$ and the more
complicated 3$d^{2}$ system V$_{2}$O$_{3}$, using the Hilbert
transform of the LDA DOS (see eq. (\ref{intg})).

\subsection{Sr$_x$Ca$_{1-x}$VO$_3$}

SrVO$_{3}$ and CaVO$_{3}$ are simple transition metal compounds
with a 3$d^{1}$ configuration. The main effect of the substitution
of Sr ions by the isovalent, but smaller, Ca ions is to decrease
the V-O-V angle from $\theta = 180^{\circ }$ in SrVO$_{3}$ to
$\theta \approx 162^{\circ }$ in the orthorhombically distorted
structure of CaVO$_{3}$. However, this rather strong bond bending
results only in a 4\% decrease of the one-particle bandwidth $W$
and thus in a correspondingly small increase of the ratio $U/W$ as
one moves from SrVO$_{3}$ to CaVO$_{3}$.\cite{Sekiyama03}

\begin{figure}[tb]
\begin{center}
\includegraphics[width=7cm]{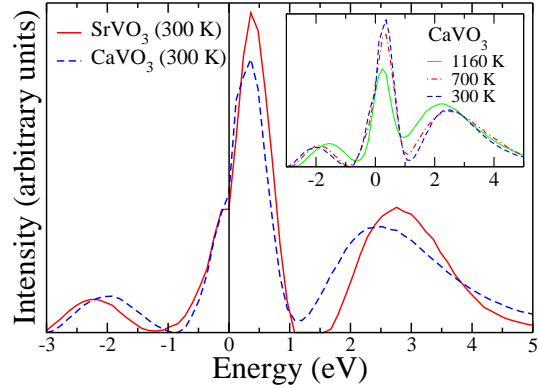}
\end{center}
\caption{LDA+DMFT(QMC) spectrum of SrVO$_{3}$ (solid line) and
CaVO$_{3}$ (dashed line) calculated at T=300 K; inset: effect of
temperature in the case of CaVO$_{3}$ [after Sekiyama \emph{et
al.}\cite{Sekiyama03}].} \label{fig_ldadmft}
\end{figure}

LDA+DMFT(QMC) spectra of SrVO$_{3}$ and CaVO$_{3}$ were calculated
by Sekiyama \textit{et al.}\cite{Sekiyama03} by starting from the
respective LDA DOS of the two materials, and are shown in Fig. 3.
These spectra show genuine correlation effects, i.e., the
formation of lower Hubbard bands at about  1.5 eV and upper
Hubbard bands at about 2.5 eV, with well pronounced quasiparticle
peaks at the Fermi energy. Therefore both SrVO$_{3}$ and
CaVO$_{3}$ are strongly correlated metals.
The DOS of the two systems shown in Fig.~\ref{fig_ldadmft} are
quite similar. In fact, SrVO$_{3}$ is slightly less correlated
than CaVO$_{3}$, in accord with their different LDA bandwidths.
The inset of Fig.~\ref{fig_ldadmft} shows that the effect of
temperature on the spectrum is small for $T \lesssim 700$~K.
Spectra of SrVO$_{3}$ and CaVO$_{3}$ were also calculated
independently by Pavarini {\em et al.}\cite{Pavarini}. The
enhancement of electronic correlations at the surface of
SrVO$_{3}$ as compared to the bulk was studied by
Liebsch\cite{Liebsch}.

Since the three $t_{2g}$ orbitals of this simple 3$d^{1}$ material
are (almost) degenerate the spectral function has the same
three--peak structure as that of the one-band Hubbard model shown
in Fig. \ref{fig:4.1}. The temperature induced decrease of the
quasiparticle peak height is also clearly seen. As noted in Sect.
3 the actual form of the spectrum no longer resembles the input
(LDA) DOS, i.e., it essentially depends only on the first three
energy moments of the LDA DOS (electron density, average energy,
band width).

\begin{figure}[tb]
\begin{center}
\includegraphics[width=8cm,angle=0]{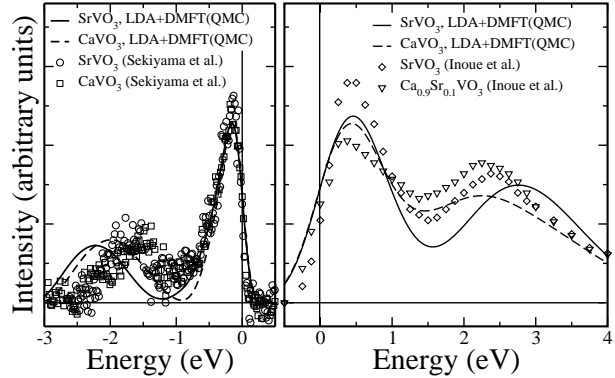}
\end{center}
\caption{Comparison of the calculated, parameter-free
LDA+DMFT(QMC) spectra of SrVO$_{3}$ (solid line) and CaVO$_{3}$
(dashed line) with experiment. Left: Bulk-sensitive
high-resolution PES (SrVO$_{3}$: circles; CaVO$_{3}$: rectangles)
[after Sekiyama \emph{et al.}\cite{Sekiyama03}]. Right: 1s XAS for
SrVO$_{3}$ (diamonds) and Ca$_{0.9}$Sr$_{0.1}$VO$_{3}$
(triangles)~\protect\cite{Inoue94}. Horizontal line: experimental
subtraction of the background intensity.
 \protect\vspace{-0.3cm}
\label{fig_XPS}}
\end{figure}

In the left panel of Fig.~\ref{fig_XPS} we compare the
LDA+DMFT(QMC) spectra at 300K to experimental high-resolution bulk
PES.
For this purpose we multiplied the theoretical spectra with the
Fermi function at the experimental temperature (20$\,$K) and Gauss
broadened with the experimental resolution of $0.1\,$eV.
\cite{Sekiyama03} The quasiparticle peaks in theory and experiment
are seen to be in very good agreement. In particular, their height
and width are almost identical for both SrVO$_{3}$ and CaVO$_{3}$.
The difference in the positions of the lower Hubbard bands may be
partly due to (i) the subtraction of the (estimated) oxygen
contribution which might also remove some $3d$ spectral weight
below $-2$~eV, and (ii) uncertainties in the {\em{ab-initio}}
calculation of $\bar{U}$. In the right panel of Fig.~\ref{fig_XPS}
we compare with XAS data of Inoue \textit{et
al.}~\cite{Inoue94,XAS}. We consider core-hole life time effects
by Lorentz broadening the spectrum with 0.2~eV~\cite{Krause79},
multiplying with the inverse Fermi function (80K), and then Gauss
broadening with the experimental resolution of
$0.36\,$eV~\cite{Inoue03}. Again, the overall agreement of the
weights and positions of the quasiparticle and upper $t_{2g}$
Hubbard band is good, including the tendencies when going from
SrVO$_{3}$ to CaVO$_{3}$ (Ca$_{0.9}$Sr$_{0.1}$VO$_{3}$ in the
experiment). For  CaVO$_{3}$ the weight of the quasiparticle peak
is somewhat lower than in the experiment. In contrast to one-band
Hubbard model calculations, our material specific results
reproduce the strong asymmetry around the Fermi energy w.r.t.
weights and bandwidths. Our results also give a different
interpretation of the XAS than in Ref.\cite{Inoue94} where the
maximum at about $2.5\,$eV was attributed to an $e_g$ band and not
to the $t_{2g}$ upper Hubbard band. The slight differences in the
quasiparticle peaks (see Fig.~\ref{fig_ldadmft}) lead to different
effective masses, namely $m^*/m_0\!=\!2.1$ for SrVO$_{3}$ and
$m^*/m_0\!=\!2.4$ for CaVO$_{3}$. These theoretical values agree
with $m^{\ast }/m_{0}\!=\!2-3$ for SrVO$_{3}$ and CaVO$_{3}$ as
obtained from de Haas-van Alphen experiments and thermodynamics
\cite{old_experiments,Inoue02}.

\subsection{V$_{\!2}$O$_{3}$}

The physical properties of ${\rm V_2O_3}$, and the metal-insulator
transition in its paramagnetic phase, have been subject of
experimental and theoretical studies for more than 30 years.
Recent advances in experimental PES and the microscopic modeling
of correlated electron systems by  the LDA+DMFT approach have led
to essential new insights into this correlation-induced
phenomenon\cite{Held01a,mo02,Keller04}.

\begin{figure}
\begin{center}
\includegraphics[width=7cm]{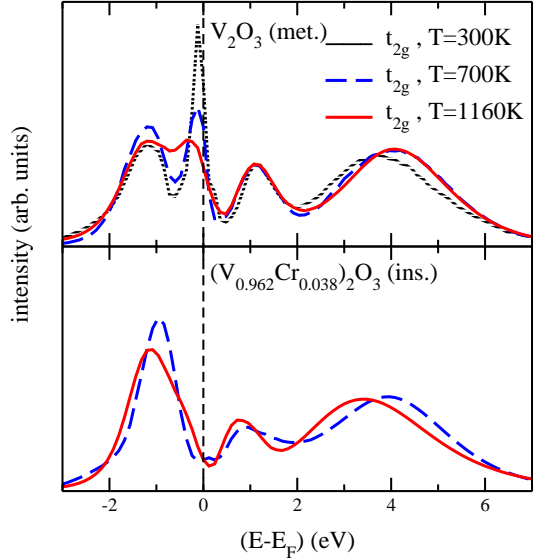}\\
\end{center}
\caption{LDA+DMFT(QMC) spectra for paramagnetic insulating ${\rm
(V_{0.962}Cr_{0.038})_2O_3}$ and metallic ${\rm V_2O_3} $ at
 $U=5$~eV [after Keller
\emph{et al.}\cite{Keller04}].} \label{fig:aw_temp}
\end{figure}

Using the crystal structure of paramagnetic metallic (PM) ${\rm
V_2O_3}$ and paramagnetic insulating (PI) ${\rm
(V_{0.962}Cr_{0.038})_2O_3}$, respectively, as input we performed
LDA+DMFT(QMC) calculations with one $a_{1g}$ and two degenerate
$e_g^\pi$ bands. To study the metal-insulator transition at
experimentally relevant temperatures we calculated at $T=700$\,K
and $T=300$\,K. Since the computational effort is proportional to
${T^{-3}}$ the low temperature calculations were computationally
very expensive. Fig.\ \ref{fig:aw_temp} shows the results of our
calculations at $T = 1160$\,K, $T=700$\,K, and $T=300$\,K for
metallic ${\rm V_2O_3}$, and at $T = 1160$\,K and $T=700$\,K for
insulating ${\rm (V_{0.962}Cr_{0.038})_2O_3}$. In the metallic
phase, the incoherent features are hardly affected by a change in
temperature, whereas the quasiparticle peak becomes sharper and
thus more pronounced at lower temperatures. This behavior also
occurs in the Anderson impurity model~\cite{Hewson93} and has its
origin in the smoothing of the Kondo-Abrikosov-Suhl resonance at
temperatures  larger than the Kondo temperature. However, in
contrast to the Anderson impurity model this smoothing occurs here
at considerably lower temperatures which is apparently an effect
of the DMFT self-consistency cycle.

Before focusing on  the comparison with experiment, let us briefly
discuss a peculiarity of the Mott-Hubbard transition which is due
to the orbital degrees of freedom.~\cite{Keller04} In the one-band
Hubbard model the Mott-Hubbard transition is characterized by the
disappearance of the quasiparticle weight $Z\rightarrow \infty$.
However, for \emph{inequivalent} $a_{1g}$ and $e_g^\pi$ orbitals
as is the case considered here only the quasiparticle weight of
the $e_g^\pi$ orbitals diverges, while that of the
 $a_{1g}$ orbital stays finite. Instead, the effective
chemical potential at low energies, i.e., $\mu - \Sigma(\mu)$ in
Eq.\  (\ref{intg}), moves outside the non-interacting LDA DOS. As
a consequence not the width but the \emph{height} of the $a_{1g}$
quasiparticle resonance goes to zero at the Mott-Hubbard
transition (albeit with a very much reduced width). For the
$e_{g}^\pi$ orbitals the transition is characterized by a combined
shrinking of width and height. In this context we note that the
pinning of the height of the spectrum at the Fermi energy -- valid
in DMFT for a single orbital\cite{MH} (or degenerate orbitals) --
does not hold in the case of inequivalent orbitals. In our case,
the volume enclosed by the $a_{1g}$ orbital shrinks at the expense
of the $e_{g}^\pi$ orbitals. Only the total volume remains
constant in accord with Luttinger's theorem.~\cite{LT}

To be able to compare with experimental photo\-emission spectra,
the LDA+DMFT results~\cite{Held01a} were multiplied with the Fermi
function at the experimental temperature ($T \approx 180$\,K) and
broadened with a $0.09$\,eV Gaussian to account for the
experimental resolution.~\cite{mo02} The same procedure was used
for the comparison with x-ray spectroscopy data (with an inverse
Fermi function at $T=300$\,K and a broadening of $0.2$\,eV taken
from experiment). On the experimental side, the PES of Refs.\
\cite{Schramme00,mo02} were corrected for the inelastic
Shirley-type background which also removes the O-$2p$
contribution. All experimental and theoretical curves were
normalized to yield the same area (which is a measure of the
occupation of the vanadium $t_{2g}$ bands).

\begin{figure}
\begin{center}
\includegraphics[width=10cm]{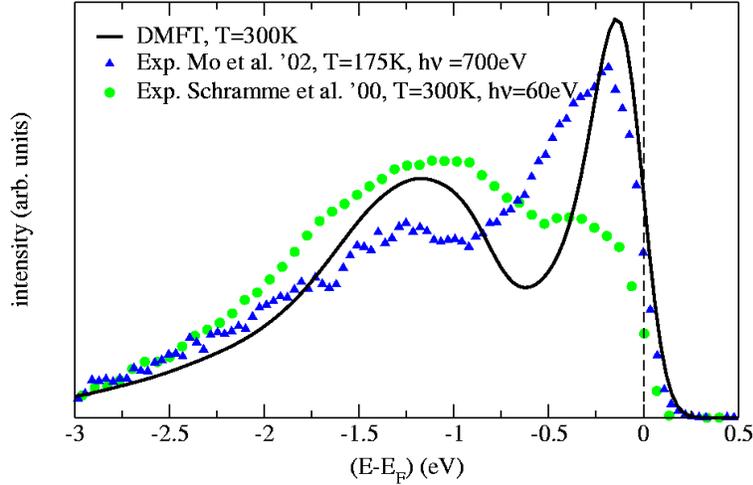}\\
\end{center}
\caption{Comparison of LDA+DMFT(QMC) results at $T=300$\,K with
photoemission data by Schramme {\em et al.}~\cite{Schramme00} and
Mo {\em et al.}~\cite{mo02} for metallic V$_2$O$_3$ [after Keller
\emph{et al.}\cite{Keller04}].} \label{fig:PES_mo}
\end{figure}

In Fig. \ref{fig:PES_mo}, the LDA+DMFT results at $300$\,K are
compared with early photoemission spectra by
Schramme~\cite{Schramme00} and recent high-resolution
bulk-sensitive photoemission spectra by Mo {\em et
al.}~\cite{mo02} The strong difference between the experimental
results is now known to be due to the distinct surface sensitivity
of the earlier data. In fact, the photoemission data by Mo {\em et
al.}~\cite{mo02} obtained at $h\nu = 700$\,eV and $T=175$\,K
exhibit, for the first time, a pronounced quasiparticle peak. This
is in good qualitative agreement with our low temperature
calculations. However, the experimental quasiparticle peak has
more spectral weight. The origin for this discrepancy, for a
system as close to a Mott transition as V$_2$O$_3$, is presently
not clear.

\begin{figure}
\begin{center}
\includegraphics[width=10cm]{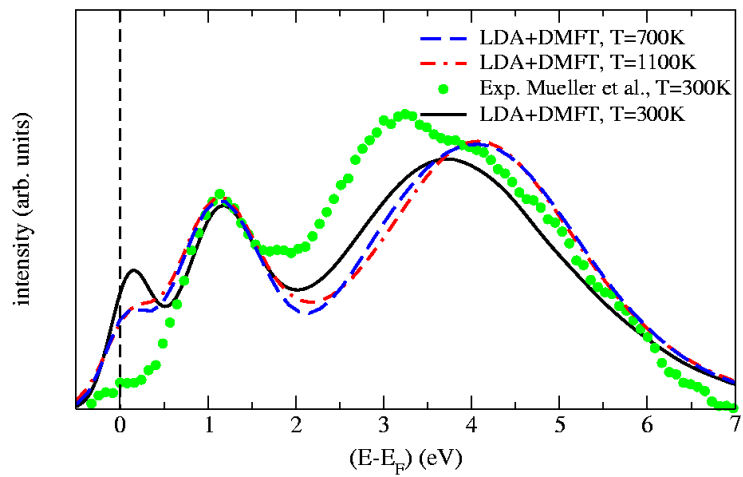}\\
\end{center}
\caption{Comparison of LDA+DMFT(QMC) results with X-ray absorption
data by M\"uller {\em et al.}~\cite{mueller97} for metallic
V$_2$O$_3$ [after Keller \emph{et al.}\cite{Keller04}].}
\label{fig:XAS_mueller}
\end{figure}

While the comparison with PES data provides important insight into
the physics of V$_2$O$_3$, more than half of the theoretical
spectrum lies above $E_F$. For this region we compare our results
at $1160$\,K, $700$\,K, and $300$\,K with O 1$s$ X-ray absorption
spectra (XAS)\cite{XAS} for V$_2$O$_3$ at $300$\,K by M\"uller
{\em et al.}~\cite{mueller97} (see Fig. \ref{fig:XAS_mueller}).
The theoretical spectra above $E_F$ are found to be almost
independent of temperature. Namely, there is a shoulder at higher
temperatures developing into a small peak at low temperatures
($300$\,K)) which is the residue of the quasiparticle peak.
Furthermore, at $1.1$\,eV there is a rather narrow peak, and at
about $4.2$\,eV a broad peak. These two peaks are parts of the
upper Hubbard band, and are due to the presence of more than one
type of correlated electron in the problem ($a_{1g}$, $e_g^\pi$)
with Hund's rule coupling $J$. The latter leads to a splitting of
the upper Hubbard band. Hence, the relative position of those two
peaks depends sensitively on the value of $J$. In particular, a
slightly smaller value of $J$ will make the agreement with
experiment even better.

\section{Volume collapse in
Cerium} \label{Ce}

Cerium (Ce) exhibits a transition from the  $\gamma$- to the
$\alpha$-phase with increasing pressure or decreasing temperature.
This transition is accompanied by an unusually large volume change
of 15\% \cite{EXPT}, much larger than the 1-2\% volume change in
${\rm V_{2}O_{3}}$.  The $\gamma$-phase may also be prepared in
metastable form at room temperature in which case the
$\gamma$-$\alpha$ transition occurs under pressure at this
temperature\cite{Olsen}. Similar volume collapse transitions are
observed under pressure in Pr and Gd (for a recent review, see
McMahan \textit{et al.}~\cite{JCAMD}).
 It is generally believed
that these transitions arise from changes in the degree of $4f$
electron correlations, as is reflected in both the Mott
transition\cite{JOHANSSON} and the Kondo volume collapse
(KVC)\cite{KVC,KVC1}  models.

The Mott transition model envisions a change from itinerant,
bonding character of the $4f$-electrons in the $\alpha$-phase to
non-bonding, localized character in the $\gamma$-phase, driven by
changes in the $4f$-$4f$ inter-site hybridization. Thus, as the
ratio of the $4f$ Coulomb interaction to the $4f$-bandwidth
increases, a Mott transition occurs to the $\gamma$-phase, similar
to the Mott-Hubbard transition of the 3$d$-electrons in ${\rm
V_{2}O_{3}}$.

The Kondo volume collapse\cite{KVC} scenario ascribes the collapse
to a strong change in the energy scale associated with the
screening of the local $4f$-moment by conduction electrons (Kondo
screening), which is accompanied by the appearance of an
Kondo-Abrikosov-Suhl-like quasiparticle peak at the Fermi level.
Indeed, the $\gamma$-$\alpha$-transition can be described using a
single-impurity model where the model parameters are determined
from DFT/LDA and spectroscopy.\cite{KVC1}

In the KVC model the change of the Ce-$4f$-electron spectrum
across the transition is, in principle, very similar to that in
the Mott scenario, i.e., there will be a strong reduction of the
spectral weight at the Fermi when going from the $\alpha$- to the
$\gamma$-phase. The subtle difference comes about by the
$\gamma$-phase having metallic $f$-spectra with a strongly
enhanced effective mass as in a heavy fermion system, in contrast
to the $f$-spectra characteristic of an insulator in the case of
the Mott scenario. The $f$-spectra in the Kondo picture also
exhibit Hubbard side-bands not only in the $\gamma$-phase, but in
the $\alpha$-phase as well, at least close to the transition.
While local-density and static mean-field theories used in the
Mott transition model up to now\cite{JOHANSSON} correctly yield
the Fermi-level peaks in the $f$-spectra for the $\alpha$-phase,
they cannot reproduce the Hubbard side-bands since these
treatments neglect genuine correlation effects. By contrast, DMFT
solutions of both Hubbard and periodic Anderson models do exhibit
the Hubbard side-bands in the $\alpha$-like regimes.\cite{Held00a}

Typically, the Hubbard model and the periodic Anderson model are
considered paradigms for the Mott and KVC model, respectively.
Although both models describe quite different physical situations
it was recently shown that they lead to a surprisingly similar
behavior at finite temperatures. Namely, for increasing Coulomb
interaction the spectrum and local magnetic moment show very
similar features. This is also the case for the phase diagram and
the charge compressibility of the periodic Anderson model with
nearest neighbor hybridization.\cite{Held00a,Held00b} From this
point of view the two scenarios are no longer really distinct, at
least at temperatures relevant for the description of the
$\alpha$-$\gamma$ transition.

For a realistic  calculation of the Ce $\alpha$-$\gamma$
transition, we employ  the full Hamiltonian calculation described
in Sect. \ref{LDA+DMFT} where the parameters entering the
one-particle Hamiltonian were calculated by LDA and the $4f$
Coulomb interaction $U$ along with the associated $4f$ site energy
shift by constrained LDA (for details of the two calculations
presented here see Refs.~\cite{JCAMD,McMahan01,McMahan03} and
Ref.~\cite{Zoelfl01}).

\begin{figure}[tb]
\begin{center}
\includegraphics[width=6cm]{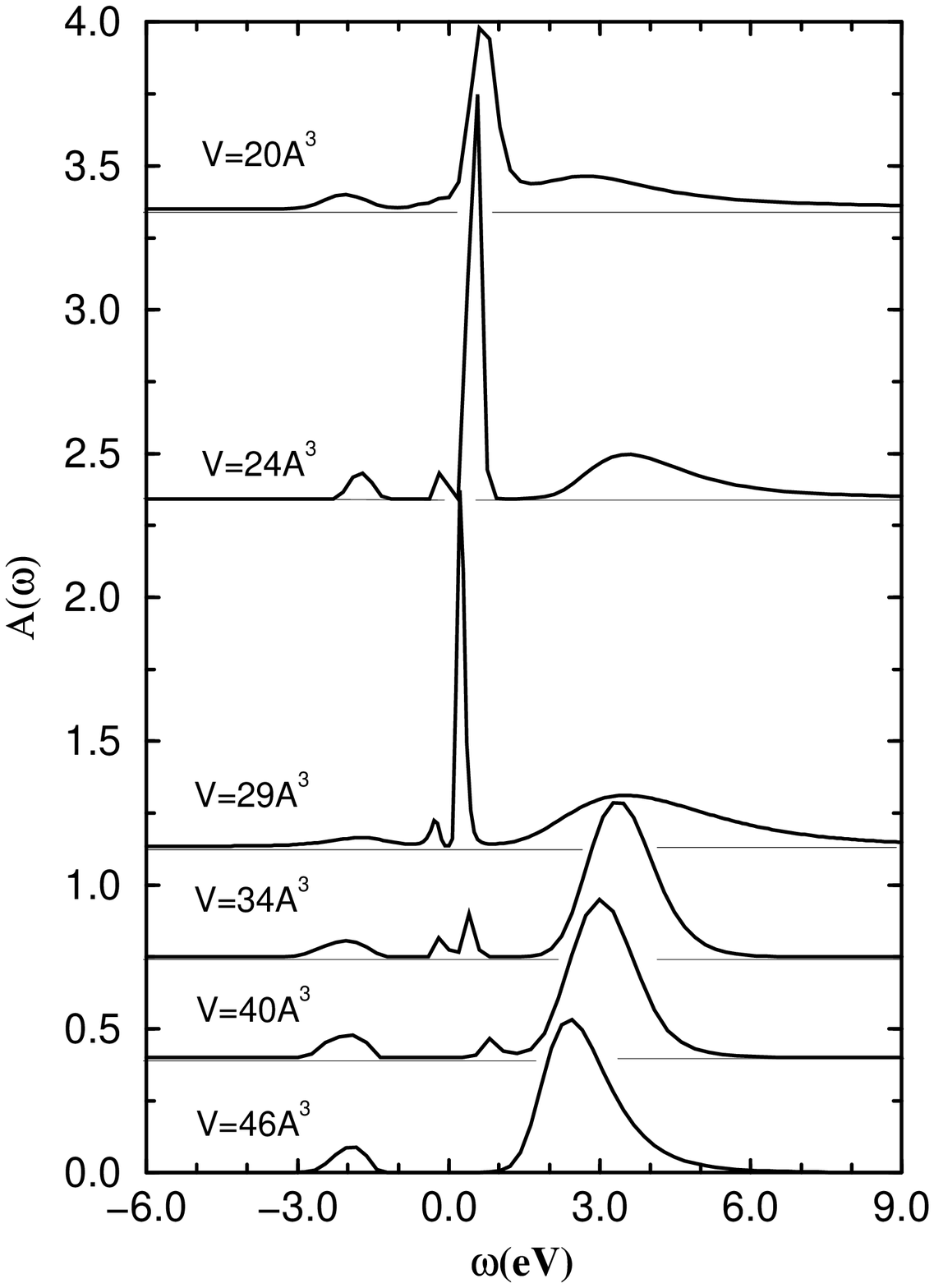}
\includegraphics[width=6cm]{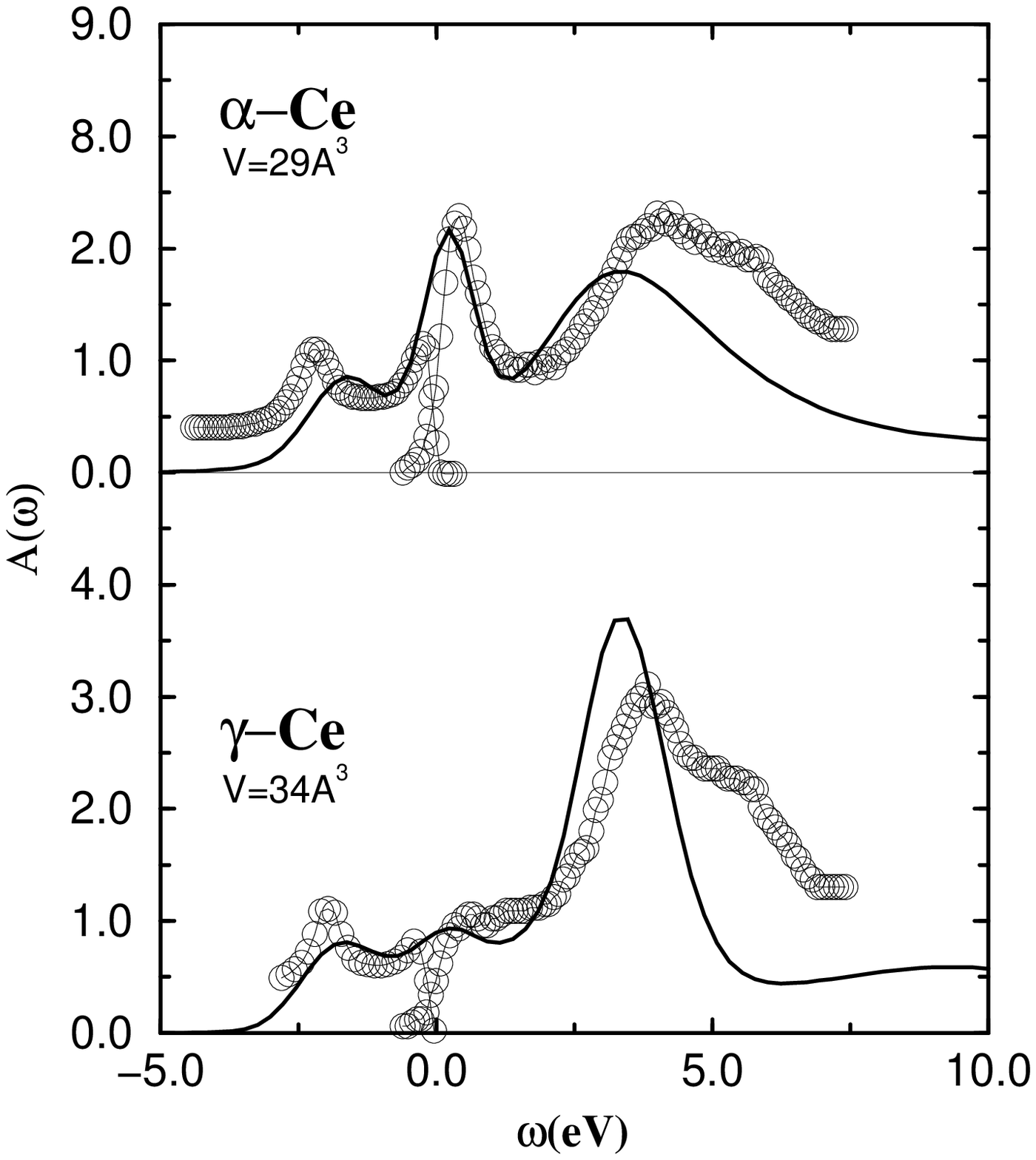}
\end{center}
\caption{Spectral functions for Ce as calculated by LDA+DMFT(QMC).
Left: $4f$ spectral function $A(\omega)$ at different volumes and
 $T\!=\!632\,$K ($\omega\!=\!0$: chemical potential;
curves are offset as indicated; $\Delta
\tau\!=\!0.11\!$$\,$eV$^{-1}$). Right: Total LDA+DMFT
$spdf$-spectrum (solid line) shown in comparison with the measured
PES\cite{PES_Wieliczka} and BIS\cite{BIS_Wuilloud} (circles) for
$\alpha$- (upper part) and $\gamma$-Ce (lower part) at $T=580\,$K
[after McMahan \emph{et al.}\cite{McMahan03}]. \label{figSpecCe}}
\end{figure}

The LDA+DMFT(QMC) spectral evolution of the Ce $4f$-electrons is
presented in the left panel of Fig.~\ref{figSpecCe}.
 At a volume per atom $V\!=\!20\,$\AA$^3$, this figure
 shows that almost the entire
spectral weight lies in a large quasiparticle peak with a center
of gravity slightly above the chemical potential.  This is similar
to the LDA solution; however, a weak upper Hubbard band is also
present even at this small volume.  At the volumes $29$$\,$\AA$^3$
and $34$$\,$\AA$^3$ which approximately bracket the
$\alpha$-$\gamma$ transition, the spectrum has a three peak
structure.
 Finally, at $V\!=\!46$$\,$$\,$\AA$^3$ the central peak has
disappeared, leaving only the lower and upper Hubbard bands. An
important  difference to ${\rm V_{2}O_{3}}$ is the metallic
feature of the $spd$-spectrum. Thus Ce remains a metal across the
transition which is monitored by a vanishing $4f$ quasiparticle
resonance.

To study the energetic changes associated with the rapid change of
the quasiparticle weight at the Fermi energy, we calculate the
DMFT energy per site for the model Hamiltonian  (\ref{Hint})
\begin{eqnarray}
E_{\rm DMFT}&\! =\!& \frac{T}{N} \sum_{n {\bf k} \sigma} {\rm
Tr}({ H}^0_{\rm LDA}({\bf k}) { G}_{\bf k}(i\omega_n))
e^{i\omega_n 0^+} + U_f \, d. \label{Eng}
\end{eqnarray}
Here, Tr denotes the trace over the $16\times16$ matrices, $T$ the
temperature, $N$ the number of ${k}$ points, $G_{\bf k}$ the Green
function matrix w.r.t.\ the orbital indices, ${ H}^0_{\rm
LDA}({\bf k})$ the LDA one-particle matrix
, and
\begin{equation}
d = \frac12 {\sum}_{m\sigma,m'\sigma'}' \langle
\hat{n}_{ifm\sigma}\, \hat{n}_{ifm'\sigma'}\rangle \label{double}
\end{equation}
is a generalization of the one-band double occupation for
multi-band models. The prime on the sum indicates that at least
two of the indices of an operator have to be different.

\begin{figure}[tb]
\begin{center}
\includegraphics[width=10cm]{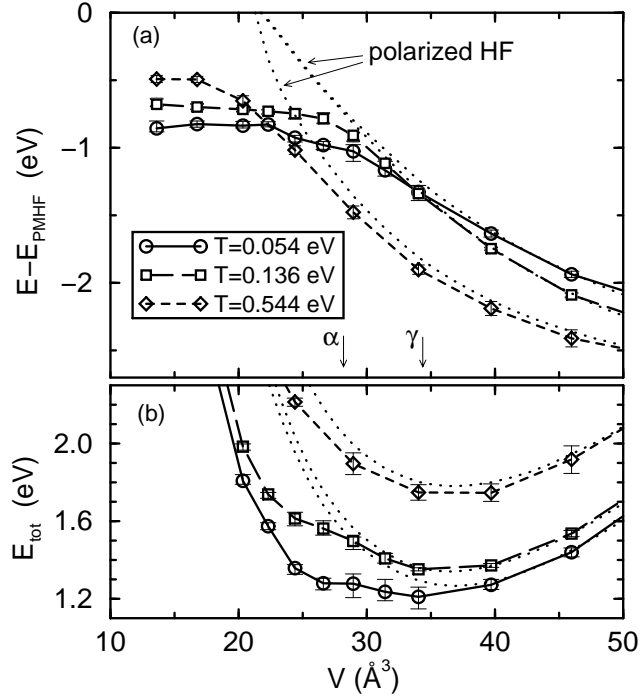}
\end{center}

\caption{(a) Correlation energy $E_{\rm DMFT}\!-\!E_{\rm PMHF}$
shown as a function of atomic volume (symbols) and Hartree-Fock
(HF) energy $E_{\rm AFHF}\!-\!E_{\rm PMHF}$ of a polarized
solution (dotted lines);
arrows: observed volume collapse from the $\alpha$- to the
$\gamma$-phase. The correlation energy is seen to bend away from
the HF energy of the  polarized solution in the region of the
transition. (b)
Total energy near $V\!=\!26$--$28$ \AA$^3$ (same symbols as in
(a));
curves at $T=0.544\,$eV were shifted downwards by $-0.5\,$eV to
match the energy range [after Held \emph{et al.}\cite{McMahan01}].
\label{figE}}
\end{figure}

Fig.~\ref{figE}a shows our calculated DMFT(QMC) energies $E_{\rm
DMFT}$ as a function of atomic volume at three temperatures  {\it
relative} to the paramagnetic Hartree-Fock (HF) energies $E_{\rm
PMHF}$ of the Hamiltonian (\ref{Hint}), i.e., the energy
contribution due to {\em electronic correlations}. We also present
the HF energies of a polarized solution which basically represent
a non-self-consistent LDA+U calculation and reproduce $E_{\rm
DMFT}$ at large volumes and low temperatures. With decreasing
volume the DMFT energies bend away from the polarized HF
solutions. Thus,  at $T\!=\!0.054\,$eV$\,\approx 600\,$K a region
of negative curvature in $E_{\rm DMFT}\!-\!E_{\rm PMHF}$ is
evident within the observed two phase region (arrows).

Fig.~\ref{figE}b presents the calculated LDA+DMFT total energy
$E_{\rm tot}(T)\!=\!E_{\rm LDA}(T)\!+\!E_{\rm DMFT}(T)\!-\!E_{\rm
mLDA}(T)$ where $E_{\rm mLDA}$ is the energy of an LDA-like
solution of the Hamiltonian (\ref{Hint}) \cite{mLDA}.  Since both
$E_{\rm LDA}$ and $E_{\rm PMHF}\!-\!E_{\rm mLDA}$ have positive
curvature throughout the volume range considered, it is the
negative curvature of the correlation energy in Fig.~\ref{figE}a
which leads to the dramatic depression of the LDA+DMFT total
energies in the range $V\!=\,$26-28$\,$\AA$^3$ for decreasing
temperature, in contrast to the smaller changes near
$V\!\!=\!34\,$\AA$^3$ in Fig.~\ref{figE}b.  This trend is
consistent with a double-well structure emerging at still lower
temperatures (which is prohibitively expensive for QMC
simulations), and with it a first-order volume collapse.
 This is in
reasonable agreement with the experimentally found volume
collapse. Other physical quantities like the free energy and the
specific heat are discussed in Ref.~\cite{McMahan03}. We note that
a similar scenario has recently been proposed by Savrasov
\textit{et al.} \cite{SAVRASOV} for the $\delta$-$\alpha$
transition in Pu on the basis of LDA+DMFT calculations. These
authors solve the DMFT equations by an ansatz inspired by iterated
perturbation theory\cite{DMFT_georges}  and include a modification
of the DFT/LDA step to account for the density changes introduced
by the DMFT.

The comparison of various static physical quantities of Ce
calculated by LDA+DMFT(QMC) and LDA+DMFT(NCA)\cite{Zoelfl01}
with experiment
shows fair to good agreement in the overall behavior and, except
for the susceptibility, even in absolute values.

Finally, in the right panel of Fig.~\ref{figSpecCe} we compare the
spectral functions
 calculated by LDA+DMFT(QMC) (see left panel of Fig.~\ref{figSpecCe})
 with experiment\cite{PES_Wieliczka}.
The main contribution to the PES of $\alpha$-Ce (right panel of
Fig.~\ref{figSpecCe}) is seen to come from the energy range
between $3\:$eV and $7\:$eV, which is attributed to $4f^2$ final
state multiplets. In the calculated spectrum all excitations to
$4f^2$ states are described by the featureless upper Hubbard band.
As a consequence of the simplified interaction model all doubly
occupied states are degenerate. This shortcoming in our
calculation is responsible for the sharply peaked main structure.
The neglected exchange interaction would produce a multiplet
structure closer to experiment. The calculated $f$-spectrum shows
a sharp quasiparticle resonance slightly above the Fermi energy,
which is the result of the formation of a singlet state between
$f$- and conduction states. We thus suggest that the spectral
weight seen in the experiment is a result of this quasiparticle
resonance. In the lower part of the right panel of
Fig.~\ref{figSpecCe}, a comparison between experiment and our
calculation for $\gamma$-Ce is shown. The most striking difference
to the result for $\alpha$-Ce (upper part of the right panel of
Fig.~\ref{figSpecCe}) is the absence of the quasiparticle
resonance in the $\gamma$-phase
 which is in agreement
with our calculations. Nonetheless $\gamma$-Ce remains metallic
with spectral weight arising from the $spd$-electrons at the Fermi
energy. Altogether, one can say that the agreement with the
experimental spectrum is very good --- comparable to the accuracy
of LDA for much simpler systems.

\section{Conclusion}
\label{conclusion}

In this paper we discussed the set-up and several applications of
the computational scheme LDA+DMFT which merges two
non-perturbative, complementary techniques for the theoretical
investigation of many-body systems in solid state physics. Using
the band structure calculated within local density approximation
(LDA) as input, the missing electronic correlations are introduced
by dynamical mean-field theory (DMFT). Thereby LDA+DMFT allows one
to perform \emph{ab initio} calculations of real materials with
strongly correlated electrons, i.e., electronic systems close to a
Mott-Hubbard metal-insulator transition, heavy fermions, and
$f$-electron materials. The physical properties of such systems
are characterized by the correlation-induced generation of small,
Kondo-like energy scales which are missing in the LDA and which
make the application of genuine many-body techniques necessary.

On a technical level LDA+DMFT requires the solution of an
effective self-consistent, multi-band Anderson impurity problem by
numerical methods, e.g., QMC. The investigation of quantum
impurity problems in the last 40 years, in particular the
development of theoretical and numerical methods to solve the
Kondo problem, is therefore a prerequisite for a successful
treatment of lattice problems within DMFT.

The application of LDA+DMFT to the transition metal oxides
Sr$_x$Ca$_{1-x}$VO$_3$, V$_{2}$O$_{3}$ as well as elemental Cerium
discussed in this paper yield, for example, spectral functions
which can be compared with spectroscopic measurements. Remarkably
good agreement was found with photoemission and x-ray absorption
data of bulk-sensitive experiments.
The spectral function of correlated metals are quite generally
characterized by a lower Hubbard band, a quasiparticle peak near
the Fermi energy, and an upper Hubbard band; the latter is split
by Hund's rule coupling in the case of more than one type of
orbital. In the upper Hubbard band the orbital structure of a
correlated material is therefore particulary clearly displayed.
This spectrum is similar to that of a localized orbital at the
impurity site hybridizing with free conduction electrons (``Kondo
problem'') although its physical origin is quite different.
Namely, in correlated bulk materials the same sort of electrons
can provide both the local moments and their screening. DMFT is
able to give a consistent explanation of this surprising
feature.

 In spite of the remarkable successes of the
LDA+DMFT approach for the investigation of correlation effects in
real materials the method still needs to be improved. Namely, at
present the LDA band structure serves only as input information
for the DMFT, but there is no feedback from DMFT to LDA. Since
correlation effects can, in principle, change the charge
distribution on which the LDA band structure depends one needs to
feed the changes computed by DMFT back into LDA, and repeat the
calculation until convergence is reached in both parts; for a
first implementation see Ref.\cite{SAVRASOV}.

Realistic calculations should not only include the orbitals of the
correlated electrons, but all orbitals. While this is already the
case in the Cerium calculations presented here, it is difficult to
do in the case of transition metal oxides. To overcome these
difficulties, an extended computational scheme has recently been
developed in Wannier basis.\cite{Wannier} It will be applied in
our future investigations to perform full orbital calculations for
correlated materials.

Besides improving the self-consistency of the LDA+DMFT method,
there are also  attempts to improve on the LDA and the DMFT part.
For example, instead of LDA the so-called GW approximation can be
employed. The main advantage is that GW is a purely diagrammatic
approach. Combined with DMFT, GW+DMFT includes the full
contribution of the Hartree diagram, the Fock diagram, and the
bubble diagrams for the screening of the Coulomb interaction as
well as the (DMFT) local contribution of all Feynman diagrams. A
first simplified implementation was reported by Biermann
\textit{et al.}\cite{Biermann}.

Furthermore, non-local correlations may be taken into account by
cluster extensions of DMFT \cite{DCA,CDMFT} (for a review see
Maier \textit{et al.}~\cite{Maier}), where instead of one Anderson
impurity one now has several coupled sites hybridizing with the
fermionic bath.

\section{Acknowledgments}

 We thank J. W. Allen, S. Horn, and S. Suga for
valuable discussions. This work was supported by the Deutsche
Forschungsgemeinschaft through Sonderforschungsbereich 484 and the
Emmy Noether program, by the Russian Basic Research Foundation
through grants RFFI-GFEN-03-02-39024\_a and RFFI-04-02-16096, by
the joint UrO-SO project N22, Grant of President of Russian
Federation for young scientists MK-95.2003.02, by the Dynasty
Foundation and International Center for Fundamental Physics in
Moscow program for young scientists 2004, and by the Russian
Science Support Foundation program for young PhD of Russian
Academy of Science 2004. We thank A. Sandvik for making his
maximum entropy code available to us. Computations were performed
at the John von Neumann Institut for Computing, J\"ulich, and the
Leibniz-Rechenzentrum, M\"unchen.


\begin{thebibliography}{99}
\bibitem{XAS} We note that the absorption measurements were taken
at the O K-edge. Therefore the spherical symmetry of the V-2p
shell in the final state remains intact. Consequently, the XAS
data can be expected to be quite similar to (still unavailable)
inverse PES data.
\bibitem{mLDA}
 We solve self-consistently for
$n_f$ using a $4f$ self-energy $\Sigma =U_f(n_f-\frac12)$, and
then remove this contribution from the eigenvalue sum to get the
kinetic energy.  The potential energy is taken to be $\frac12 U_f
n_f(n_f-1)$.
\end{thebibliography}
\end{document}